\documentclass{article}
\usepackage{booktabs}
\usepackage{graphics}
\usepackage{graphicx}
\usepackage{listings}
\usepackage{nameref}
\usepackage{algorithm}
\usepackage{algpseudocode}
\algrenewcommand\algorithmicrequire{\textbf{Input:}}
\algrenewcommand\algorithmicensure{\textbf{Output:}}
\usepackage{bm}
\usepackage{amsmath}
\usepackage{amssymb}
\usepackage{caption} 
\usepackage{multirow}
\usepackage{upgreek}
\usepackage{color}
\usepackage[table]{xcolor}

\usepackage{arxiv}
\usepackage[utf8]{inputenc} 
\usepackage[T1]{fontenc}    
\usepackage{hyperref}       
\usepackage{url}            
\usepackage{booktabs}       
\usepackage{amsfonts}       
\usepackage{nicefrac}       
\usepackage{microtype}      
\usepackage{lipsum}
\usepackage{graphicx}
\graphicspath{ {./images/} }
\begin{document}
\title{TPMM: Three-component Posterior Mixture Model Enables Robust Inverton Detection in Low-Depth Metagenomes and Suggests Potential Viral Invertons}
\author{
  Yi Lu \\
  Dept. of Electrical Engineering\\
  City University of Hong Kong\\
  Kowloon, Hong Kong SAR, China\\
  \And
  Jiaojiao Guan \\
  Dept. of Electrical Engineering\\
  City University of Hong Kong\\
  Kowloon, Hong Kong SAR, China\\
  \And
  Yang Shen \\
  Dept. of Electrical Engineering\\
  City University of Hong Kong\\
  Kowloon, Hong Kong SAR, China\\
  \And
  Jiayu Shang \\
  Dept. of Information Engineering\\
  Chinese University of Hong Kong\\
  New Territory, Hong Kong SAR, China\\
  \And
  Yanni Sun \\
  Dept. of Electrical Engineering\\
  City University of Hong Kong\\
  Kowloon, Hong Kong SAR, China
}
\maketitle
\begin{abstract}
Bacterial phase variation enables reversible, locus-specific phenotypic switching, often driven by DNA inversion (invertons). To identify these events, researchers commonly rely on sequencing reads that provide orientation-specific support. Metagenomic sequencing, which captures total genetic material independent of cultivation, offers a powerful platform for the comprehensive study of invertons. However, computational inverton calling from metagenomic data is difficult at low sequencing depth: hard read-support cutoffs can miss true events, while sequence-only predictors lack read-backed interpretability and uncertainty quantification. To address this, we present TPMM, a three-component posterior mixture model for inverton calling in metagenomic data. TPMM explicitly incorporates sequencing depth to formulate inverton detection as a probabilistic mixture problem. Starting from candidates flanked by inverted repeats, the model classifies the candidates into noise, low-probability, or high-probability inversion signals using read evidence. Finally, TPMM assigns posterior probabilities as soft labels and applies cumulative Bayesian False Discovery Rate control to robustly identify true invertons. On two real gut metagenomic datasets, TPMM agrees well with PhaseFinder at high depth but recovers substantially more invertons under systematic downsampling, demonstrating superior performance in sparse-data regimes. We further examine potential reversible inversion elements in viral genomes and provide supporting analyses, suggesting a broader scope for inversion-mediated regulation.

\textbf{Availability:} The source code of TPMM is available via: \href{https://github.com/KennyxxD/TPMM}{https://github.com/KennyxxD/TPMM}.

\textbf{Contact:} \href{yannisun@cityu.edu.hk}{yannisun@cityu.edu.hk}
\end{abstract}
\section{Introduction}
Bacterial phase variation is a widespread strategy that enables frequent, reversible switching of genetic states at specific loci \cite{https://doi.org/10.1046/j.1365-2958.1999.01555.x, doi:10.1128/cmr.17.3.581-611.2004, TRZILOVA202159}, thereby generating phenotypic heterogeneity even within clonal populations. In contrast to adaptation driven by point mutations, phase variation typically restricts variability to a small set of hypermutable loci. This localization allows populations to rapidly sample alternative phenotypes while limiting the fitness costs associated with an elevated mutation rate across the rest of the genome. Phase variation is therefore thought to play a central role in pathogen infection and immune evasion, and in the long-term persistence of commensals under fluctuating selective pressures, including host immunity, antibiotic exposure and spatial heterogeneity \cite{doi:10.1073/pnas.0804220105, Porter2020}. Among several molecular mechanisms of phase variation, reversible DNA inversion is particularly prominent \cite{TRZILOVA202159, 10.1093/nar/gkaa824}. This process is typically mediated by site-specific recombinases (invertases) recognizing flanking inverted repeats (IRs) to catalyze segment inversion \cite{doi:10.1128/jb.01362-06, Jin2025}. Compared with nucleotide substitutions or indels, inversion tends to occur at higher rates and is reversible, enabling stable coexistence of subpopulations with distinct phenotypic states. Functionally, a common regulatory architecture is the invertible promoter: in the ON orientation, the promoter drives transcription of downstream genes or operons, whereas in the OFF orientation transcription is inhibited \cite{Yan2022}. For example, several capsular polysaccharide loci in \textit{Bacteroides fragilis} are regulated by invertible promoters that flip between ON and OFF orientations, thereby turning expression of the corresponding biosynthesis operons on or off \cite{Troy2010BfragilisPromoterOrientations}. Invertible segments can also include alternative regulatory elements such as terminators, altering transcriptional output via additional mechanisms. Beyond intergenic invertons, gene-intersecting inversions may generate alternative protein isoforms or modify molecular specificity \cite{GRUNDY1984296, doi:10.1128/jb.160.1.299-303.1984}. Collectively, inverton-mediated regulation can act both as an expression switch and as a mechanism shaping microbial structural and functional interactions with their environment and hosts. \par
In metagenomic data, active inversion is evidenced by sequencing reads supporting both orientations of a locus \cite{Troy2010BfragilisPromoterOrientations}. Accordingly, the numbers of reads supporting the forward versus reverse orientation, and their ratio, are commonly used features for inverton detection and quantification. 
PhaseFinder, the first tool to exploit this, identifies invertons by quantifying read support for in silico generated orientations \cite{doi:10.1126/science.aau5238}. However, PhaseFinder relies on fixed thresholds, which often excludes true invertons in low-coverage regions. To address this limitation, a second class of methods primarily relies on sequence signals. DeepInverton \cite{10.1093/nar/gkae966} is designed based on the inverton sequence rather than read evidence. 
Moreover, by operating independently of reads, DeepInverton lacks direct read-backed interpretability and principled treatment of sampling uncertainty.\par
\begin{figure*}[!t]
  \centering
  \includegraphics[width=1.0\textwidth]{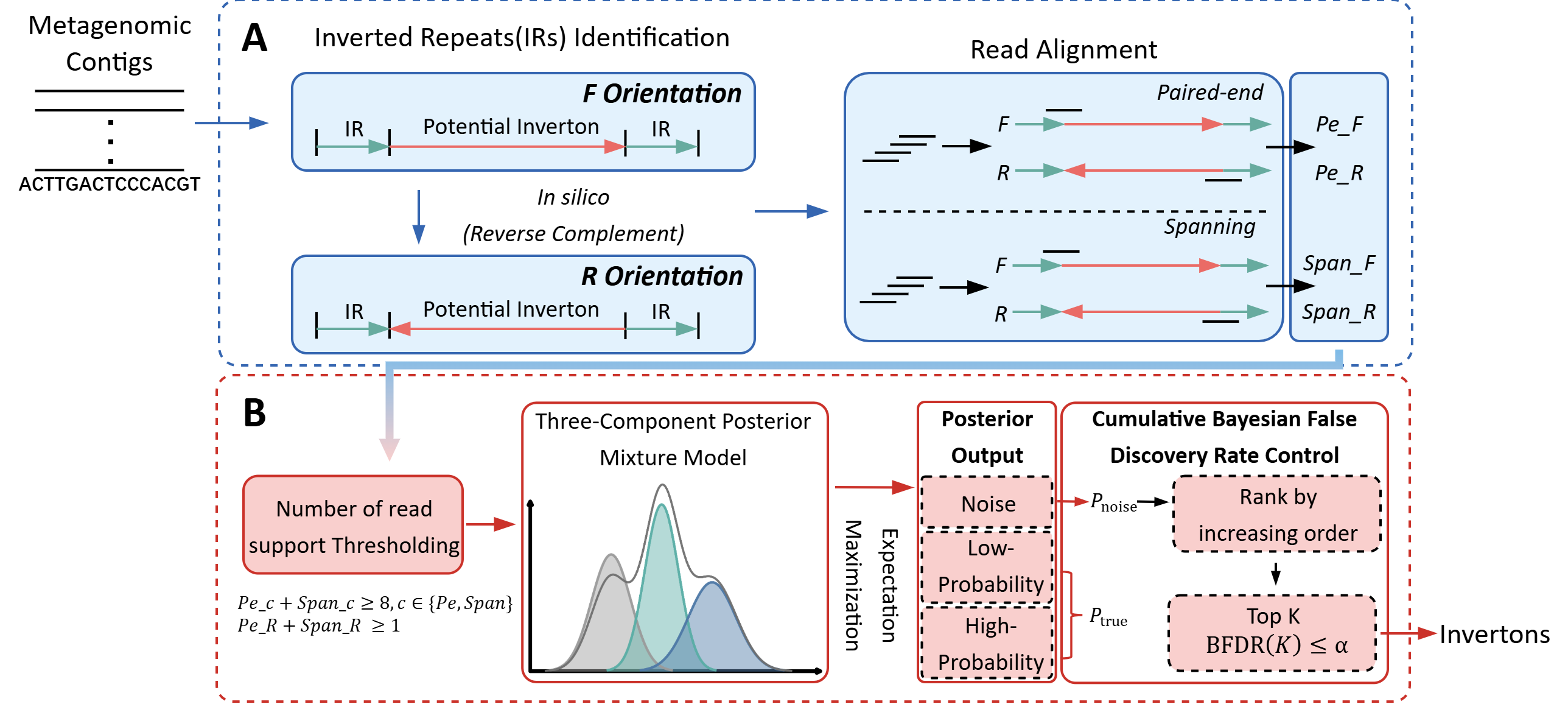}
  \caption{Overview of the pipeline. (A) Acquisition of read support evidence. PhaseFinder is utilized to quantify read counts supporting both the forward and reverse orientations. (B) Statistical inference and false discovery control. After initial thresholding of read evidence, the data is analyzed using a Three-component Posterior Mixture Model (TPMM) to estimate posterior probabilities for each locus via the Expectation-Maximization(EM) algorithm. Candidate invertons are ranked in ascending order of their posterior probability of being noise ($P_{\text{noise}}$) to calculate the cumulative Bayesian False Discovery Rate (BFDR). The final set of invertons is identified by selecting the top $K$ candidates such that $\text{BFDR}(K) \leq \alpha$, where $\alpha$ is a user-defined threshold.}
  \label{Fig1}
\end{figure*}
Here, we develop TPMM, a three-component posterior mixture model for probabilistic inverton calling. TPMM operates on the full set of candidate loci generated by PhaseFinder prior to its default filtering, specifically retaining loci with at least one supporting reverse read (via paired-end or spanning methods) and a total read count exceeding a fixed threshold. We propose an explicit generative framework for read support evidence. decomposing the candidate loci into a three-component posterior mixture representing noise, low-probability, and high-probability signals. The noise component captures spurious support for reverse orientation reads, while the low-probability and high-probability components represent true inversions occurring at low and high probabilities in the environment. The model outputs posterior probabilities as soft labels, allowing for the quantification of evidence strength and uncertainty without relying on an arbitrary binary cutoff. To identify the final set of true invertons for downstream analysis, we further control cumulative Bayesian False Discovery Rate on the posterior outputs, converting the soft labels into final hard calls \cite{Newton2004SemiparametricHierarchicalMixture}. \par
We validated TPMM on two real metagenomic datasets, including a human gut dataset and a rat gut dataset. At high read depth, TPMM shows strong agreement with PhaseFinder, indicating that it preserves conclusions when evidence is sufficient. Based on this, we performed downsampling on both datasets at three different ratios to emulate progressively lower read depth and prove that TPMM recovers substantially more invertons than PhaseFinder, demonstrating improved recovery under low-depth conditions. Moreover, the invertons identified by TPMM are significantly enriched in downstream genes, underscoring their biological significance. Finally, we proposed that viral genomes may also contain invertons or related reversible inversion elements, provide corresponding analyses and supporting evidence, extending the potential scope of inversion mediated regulation to viral systems.
\section{Methodology}\label{sec:methodology}
\subsection{Overview and Data Representation}

As illustrated in Figure \ref{Fig1}A, PhaseFinder initiates by using EMBOSS einverted to locate inverted repeats. It then generates an augmented reference containing both forward and reverse sequences \textit{in silico} for each candidate for metagenomic read alignment. Alignment evidence is derived from Paired-end (Pe) and Spanning (Span) methods, which are classified into Forward (F) or Reverse (R) orientations. A significant accumulation of R reads relative to total depth indicates the presence of an active DNA inversion. However, the default PhaseFinder algorithm relies on hard-filtering thresholds to identify positive calls, specifically requiring at least 5 paired-end reverse reads ($Pe\_R \ge 5$), 3 spanning reverse reads ($Span\_R \ge 3$), and a reverse orientation frequency of at least 1\%.
Although effective for high-coverage isolate genomes, this approach lacks the flexibility required for metagenomic data. Given the variable sequencing depth inherent in complex microbial communities, the reliance on hard thresholds inevitably leads to the omission of invertons under low-depth metagenomic data.\par
Hence, we propose TPMM, the Three-component Posterior Mixture Model. Instead of applying a rigid cutoff, TPMM evaluates read support to generate a probabilistic soft label. This strategy allows the model to adapt dynamically to varying read counts, ensuring the recovery of valid biological signals even when read support is scarce. The workflow of TPMM is shown in Figure \ref{Fig1}B. It defines the input data $\boldsymbol{x}_i$ for a candidate locus \(i\) as the raw results of read support:
\begin{equation}
     \boldsymbol{x}_i = [F_i^{\text{Pe}}, R_i^{\text{Pe}}, F_i^{\text{Span}}, R_i^{\text{Span}}],
\end{equation}
where \(F\) and \(R\) denote the number of reads supporting the forward and reverse orientations, respectively. We define the total sequencing depth for candidate \(i\) within a specific channel \(c \in \{\text{Pe, Span}\}\) as \(N_i^c = F_i^c + R_i^c\).
We initially process each metagenomic sample independently using its corresponding sequencing reads. For each sample, we apply a relaxed pre-filtering step to retain potential low-probability signals, keeping candidates with a total number of reads ($N_i^{\text{Pe}} + N_i^{\text{Span}} \ge 8$) and at least one supporting reverse read in either channel ($R_i^{\text{Pe}} + R_i^{\text{Span}} \ge 1$). The filtered candidates from all samples in the same environment (i.e., human gut or rat gut) are then collected together and processed using the framework below.
\subsection{Three-component Posterior Mixture Model (TPMM)}
We assume that the observed read counts arise from a mixture of three distinct latent biological states, denoted by \(z_i \in \{0, 1, 2\}\):
\begin{itemize}
    \item \textbf{Noise (\(k=0\)):} Background sequencing errors where no inversion exists.
    \item \textbf{Low-probability (\(k=1\)):} Rare inversion events present in a small fraction of the population.
    \item \textbf{High-probability (\(k=2\)):} Common inversion events in this environment.
\end{itemize}

The marginal likelihood for the observed data \(\boldsymbol{x}_i\) is a weighted sum of the conditional likelihoods for each component:
\begin{equation}
    P(\boldsymbol{x}_i \mid \Theta) = \sum_{k=0}^{2} \pi_k \, P(\boldsymbol{x}_i \mid z_i=k, \Theta),
\end{equation}
where \(\pi_k = P(z_i=k)\) represents the mixing proportion of state \(k\), subject to \(\sum \pi_k = 1\). The  set of model parameters is defined as:
\begin{equation}
    \Theta = \{\pi_0, \pi_1, \pi_2, \theta_1^{\text{Pe}}, \theta_1^{\text{Span}}, \theta_2^{\text{Pe}}, \theta_2^{\text{Span}}, \beta_0, \beta_1\},
\end{equation}
which includes the mixing proportions, the signal inversion rates (\(\theta\)), and the noise regression coefficients (\(\beta\)). These parameters are described in detail below.
\subsubsection{Component-wise Likelihood}
We model the number of reverse reads \(R_i^c\) in channel \(c\) as a Binomial random variable, conditioned on the total depth \(N_i^c\) and an inversion probability \(p_k^c\). Assuming conditional independence between the Pe and Span channels, the likelihood for candidate \(i\) given state \(k\) is:
\begin{equation}
    P(\boldsymbol{x}_i \mid z_i=k, \Theta) = \prod_{c \in \{\text{Pe, Span}\}} \text{Binomial}\left(R_i^c \mid N_i^c, p_k^c(i)\right).
\end{equation}
Here, \(p_k^c(i)\) represents the expected proportion of reverse reads for candidate \(i\) in state \(k\). We define these probabilities differently for signal and noise components.
For true invertons, we estimate a constant inversion rate for each channel. Thus, \(p_1^c(i) = \theta_1^c\) and \(p_2^c(i) = \theta_2^c\). To ensure model identifiability and consistent interpretation, we enforce the constraint that the ``High-probability'' component must have a higher mean inversion rate than the ``Low-probability'' component:
\begin{equation}
    \frac{\theta_2^{\text{Pe}} + \theta_2^{\text{Span}}}{2} \ge \frac{\theta_1^{\text{Pe}} + \theta_1^{\text{Span}}}{2}.
\end{equation}
Sequencing error rates often fluctuate with sequencing depth. To capture this, we model the noise parameter \(p_0(i)\) using a logistic regression function dependent on the total read depth \(N_i = N_i^{\text{Pe}} + N_i^{\text{Span}}\). To reduce nuisance parameters, we share this noise rate across both channels:
\begin{equation}
    p_0^{\text{Pe}}(i) = p_0^{\text{Span}}(i) = \sigma\left(\beta_0 + \beta_1 \ln(1 + N_i)\right),
\end{equation}
where \(\sigma(t) = (1 + e^{-t})^{-1}\) is the sigmoid function, and \(\beta_0, \beta_1\) are regression coefficients estimated from the data.
\subsubsection{Estimation and Inference}
We estimate the parameter set \(\Theta\) via the Expectation-Maximization (EM) algorithm by maximizing the total log-likelihood \(\mathcal{L}(\Theta) = \sum_{i} \ln P(\boldsymbol{x}_i \mid \Theta)\). The detailed procedure is provided in Supplementary Section 1.
After convergence, we compute the posterior probability that candidate \(i\) belongs to state \(k\), denoted as \(\gamma_{ik} = P(z_i=k \mid \boldsymbol{x}_i)\). We define the probability that a candidate is a true inverton (\(P_{\text{true}}\)) as the sum of the posterior probabilities for the low- and high-probability signal components:
\begin{equation}
    P_{\text{true}}^{(i)} = 1 - \gamma_{i0} = \gamma_{i1} + \gamma_{i2}.
\end{equation}
To determine the final set of positive calls, we employ cumulative Bayesian False Discovery Rate (BFDR) approach. We define the local false discovery rate (lfdr) for candidate \(i\) as its noise probability (\(\text{lfdr}_i = \gamma_{i0}\)). Candidates are ranked in increasing order of $\gamma_{i0}$. For a cutoff at rank \(K\), the cumulative BFDR is calculated as:
\begin{equation}
    \text{BFDR}(K) = \frac{1}{K} \sum_{j=1}^{K} \text{lfdr}_{j} =\frac{1}{K} \sum_{j=1}^{K} \gamma_{j0}.
\end{equation}
We select the largest \(K\) such that \(\text{BFDR}(K) \le \alpha\) (where \(\alpha\) is the desired error rate), providing a principled method to control false positives without relying on arbitrary read count thresholds.
\section{Results}
We analyzed two metagenomic datasets: a rat gut study \cite{WEN2025122108} and a human gut study \cite{Lewis2015CrohnsMicrobiomeStressors}. We then compared our results with PhaseFinder and DeepInverton. The rat dataset, collected from diverse habitats in Hong Kong, represents complex environmental stressors that may drive inverton turnover. The human dataset comprises 26 samples from healthy controls and 343 from individuals with Crohn’s disease, spanning multiple therapeutic regimens and longitudinal timepoints. This design captures dynamic perturbations in the gut environment, thereby increasing the likelihood of detecting inverton events. \par
Raw sequencing reads underwent quality control using Trimmomatic (v0.39) to remove adapters and low-quality bases. The parameters are provided in Supplementary Table S2. To minimize host contamination, filtered reads were aligned to the corresponding host reference genomes using Bowtie2 (v2.3.5.1), and aligned reads were discarded. The remaining high-quality, non-host reads were assembled de novo into contigs using MEGAHIT (v1.2.9) with default settings. Following preprocessing, 88 rat and 365 human gut metagenomes were retained for downstream analysis with approximately $2.81 \times 10^9$ and $2.95 \times 10^9$ reads, respectively. \par
First, we characterized the identified invertons by examining their supporting read distributions, demonstrating how the probabilistic model stratifies candidates based on evidence strength rather than arbitrary cutoffs. Second, we benchmarked our results against the standard output of the default PhaseFinder algorithm to evaluate concordance, specifically analyzing the read evidence profiles underlying the unique detections identified by each method. Next, to evaluate the impact of sequencing depth on detection reliability, we conducted a downsampling experiment, quantifying the recovery of high-confidence invertons across systematically reduced coverage levels. Subsequently, to assess functional significance, we investigated the genomic context of the identified invertons, specifically testing for enrichment near genes using a contig-preserving permutation framework that controls for local gene density. \par
Finally, we presented an interesting discovery emerging from the analysis of the human gut metagenomic data, where we observed that a subset of the identified invertons resides on viral genomes. We analyzed these potential invertons by integrating viral contig identification, genome completeness assessment, and coding sequence annotation.
\begin{figure}[H]
  \centering
  \includegraphics[width=0.65\textwidth]{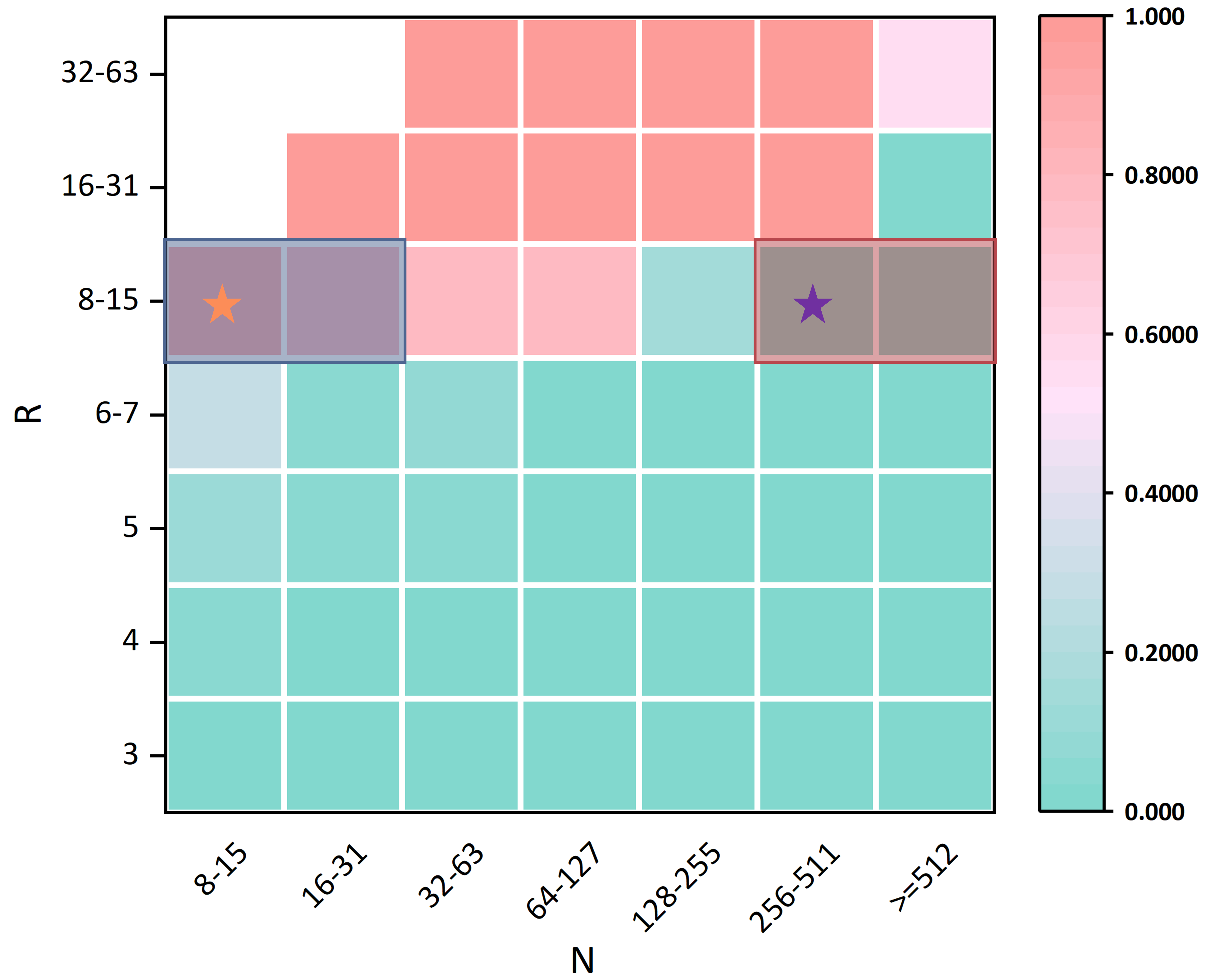}
  \caption{Heatmap of read evidence and posterior for selected samples from the human gut dataset. For each bin, we plot the median $P_{\mathrm{true}}$ across candidates falling into that bin. Bins with no observations are left blank. The yellow star indicates the bin with the largest positive enrichment (i.e., where TPMM yields the greatest excess of unique positives), whereas the purple star indicates the bin with the largest negative enrichment (i.e., where PhaseFinder yields the greatest excess of unique positives).}
  \label{Fig2}
\end{figure}
\subsection{TPMM helps define a high-confidence inverton set}
\label{3.1}
\subsubsection{Output results and comparison with PhaseFinder}
We applied TPMM to each candidate to infer its latent component membership and obtained posterior probabilities for the three mixture components. In total, we identified 3,517 invertons in the human gut dataset and 764 in the rat gut dataset. To examine whether the estimated results behave consistently with read evidence, we summarized the empirical distributions of $N_i$ and $R_i$ and constructed bins to ensure adequate coverage across the $(N_i, R_i)$ grid. We then visualized the binned relationship between $P_{\mathrm{true}}$ and read evidence in Figure \ref{Fig2}. As shown in Figure~\ref{Fig2}, for a fixed $N_i$ bin, $P_{\mathrm{true}}$ increases with $R_i$, reflecting stronger support for the inversion. Conversely, for a fixed $R_i$ bin, $P_{\mathrm{true}}$ decreases as $N_i$ increases (equivalently, as $\frac{R_i}{N_i}$ decreases). 
For downstream analyses, we controlled the cumulative BFDR at \(\alpha=0.05\) and obtained positive calls. The detailed calculation procedure, as well as an additional consistency check between \(P_{\mathrm{true}}\) and \(\mathrm{BFDR}(K)\), is provided in Supplementary Figure S1.
Then we compared the TPMM positive calls with PhaseFinder results. Here, we mainly focused on the calls uniquely identified by these two methods. The blue box in Figure \ref{Fig2} highlights TPMM-unique positives, which concentrate in bins with smaller $N_i$ but relatively higher $\frac{R_i}{N_i}$. To investigate the underlying cause, we examined the detailed read distribution and found that these discrepancies are primarily driven by insufficient absolute read support in one of PhaseFinder’s evidence channels (e.g., paired-end or spanning) due to its minimum supporting-read thresholds. In contrast, the red box marks PhaseFinder-unique calls, which are enriched at larger $N_i$ but lower $\frac{R_i}{N_i}$. To illustrate the most pronounced disagreements between two methods, we calculated the net enrichment of TPMM-unique calls over PhaseFinder-unique calls. Moreover, we provided Venn diagrams for both datasets in Supplementary Figure S2. For downstream analysis, we defined a conservative reference set as the intersection of TPMM and PhaseFinder calls at full depth, ensuring that subsequent sensitivity comparisons are anchored on events supported by both approaches. \par
\subsubsection{Comparison with DeepInverton}
In addition, we applied DeepInverton to both datasets with default parameters the authors provide. In the human gut dataset, the tool identified 1310 candidate invertons, yet only 12 possess reverse read support. Similarly, in the rat gut dataset, 4103 invertons were predicted, none of which were supported by reverse reads. This disconnection between prediction and physical evidence suggests that this tool, which relies on DNA sequences rather than alignment data, lacks the reliability and biological interpretability necessary to accurately identify active DNA inversions.
\subsection{TPMM retains higher inverton recovery under downsampling}
\begin{figure}[htbp]
  \centering
  \includegraphics[width=0.65\textwidth]{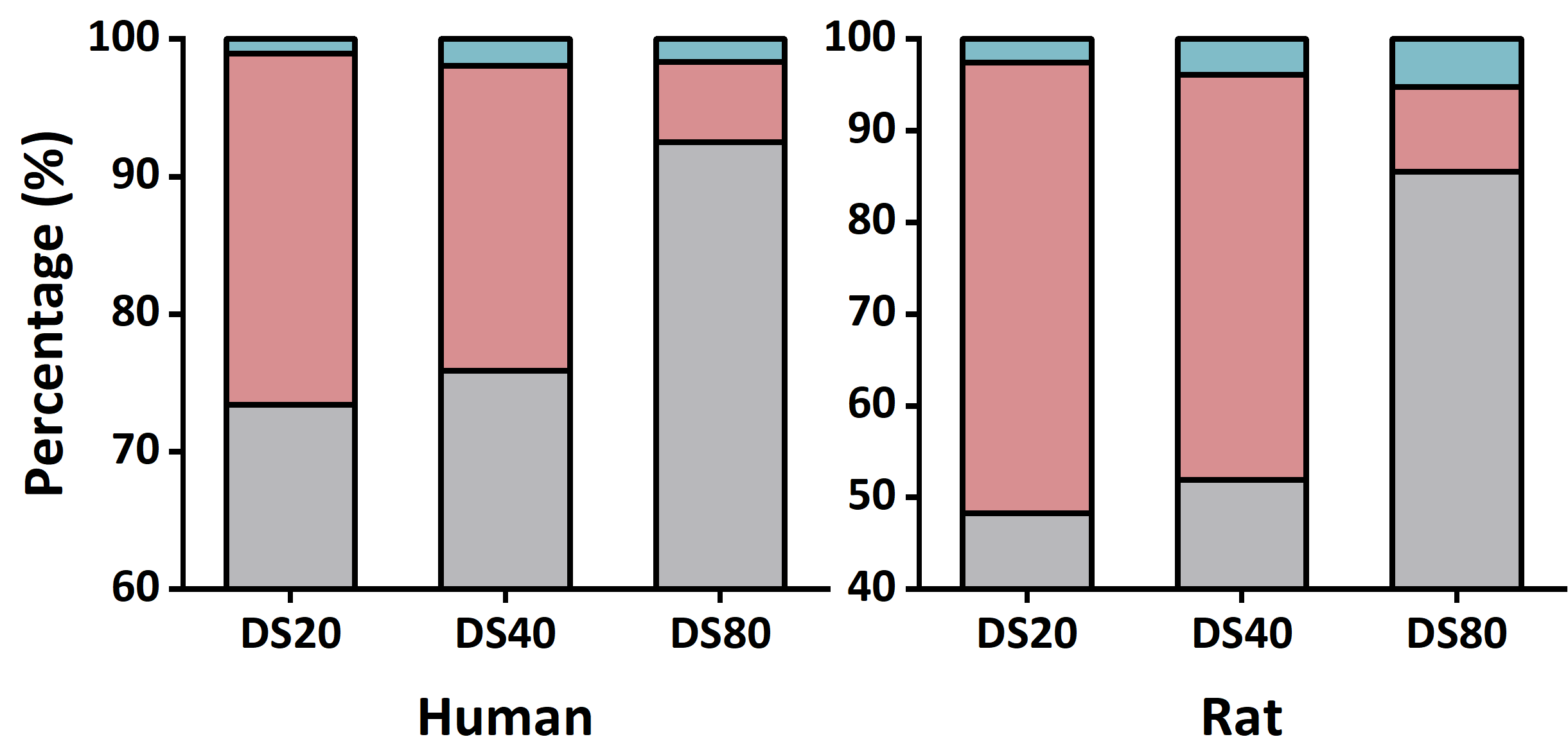}
  \caption{Composition of \(I_{\mathrm{ref}}\) calls recovered under downsampling. DS20, DS40, and DS80denote downsampled datasets retaining 20\%, 40\%, and 80\% of the original reads. Left: Human dataset; Right: Rat dataset. Stacked bars represent the proportion of \(I_{\mathrm{ref}}\) identified by both methods (common; grey), detected only by TPMM (red), and detected only by PhaseFinder (blue).}
  \label{Fig4}
\end{figure}
As mentioned above, most of invertons detected by DeepInverton in both datasets lack reverse read support. Therefore, we only compared TPMM and PhaseFinder in this experiment. To assess robustness under low-depth sequencing conditions, we performed a downsampling analysis on both datasets. For each sample, we randomly subsampled 20\%, 40\%, and 80\% of the original reads (denoted as DS20, DS40, and DS80), respectively. Subsequently, TPMM and PhaseFinder were run independently on each of the six downsampled datasets. \par
Then, we defined the high-confidence set described in previous section as $I_\mathrm{ref}$, representing the intersection of invertons identified by both methods on the full-coverage datasets: $I_\mathrm{ref} = I_\mathrm{TPMM}^\mathrm{Full} \cap I_\mathrm{PhaseFinder}^\mathrm{Full}$.
For each down-sampled condition, we mapped the identified invertons to $I_\mathrm{ref}$ based on genomic coordinates. We then quantified: (i) the fraction of $I_\mathrm{ref}$ recovered by both methods, and (ii) the proportion of $I_\mathrm{ref}$ recovered exclusively by either TPMM or PhaseFinder. As illustrated in Figure \ref{Fig4}, the subset of invertons recovered uniquely by TPMM (red bar) was consistently larger than that recovered uniquely by PhaseFinder (blue bar) across both datasets, indicating that TPMM maintains higher sensitivity under reduced sequencing depth. 

Specifically, in the human gut dataset, TPMM uniquely recovered 25.54\%, 22.15\%, and 5.80\% of the reference set under DS20, DS40, and DS80, respectively, whereas PhaseFinder uniquely recovered only 1.07\%, 1.96\%, and 1.68\%. Similarly, in the rat gut dataset, TPMM uniquely recovered 49.15\%, 44.14\%, and 9.21\%, compared to 2.54\%, 3.90\%, and 5.26\% for PhaseFinder. Notably, this performance advantage become more pronounced as sequencing depth decreased, demonstrating that TPMM is more robust in preserving the detection of reference invertons when read depth is severely limited. The absolute numbers of invertons detected under each condition are summarized in Table \ref{table1}.
\begin{table}[htbp]
\centering
\small
\begin{tabular}{ccccc}
\toprule
\multirow{2}{*}{Subset} & \multicolumn{2}{c}{TPMM-only} & \multicolumn{2}{c}{PhaseFinder-only}\\
\cmidrule(lr){2-3} \cmidrule(lr){4-5}
 & Human & Rat & Human & Rat\\
\midrule
DS20 & 262 & 58 & 11 & 3 \\
DS40 & 395 & 147& 35 & 13\\
DS80 & 145 & 124 & 42 & 78\\
\bottomrule
\end{tabular}
\caption{Unique invertons recovered under downsampling by TPMM and PhaseFinder in $I_\mathrm{ref}$  at each downsampling level.}
\label{table1}
\end{table}
\subsection{Enrichment of downstream genes near invertons}
Previous studies have reported that invertons can modulate downstream gene expression \cite{doi:10.1073/pnas.0404172101, doi:10.1073/pnas.1005039107, doi:10.1128/mbio.01339-15}. Consequently, we investigated whether the invertons identified by TPMM in two datasets are located upstream of genes more frequently than expected by chance. Detailed definitions of downstream regions are provided in Supplementary Section 2. We first performed de novo gene prediction on all contigs using Prodigal \cite{Hyatt2010} to define coding sequence (CDS) start coordinates. We then quantified the proximity of invertons to downstream CDS start sites and assessed statistical significance using a contig-preserving permutation framework that accounts for contig structure and local gene density. \par
Briefly, for a given window size $w$, we determined whether at least one annotated CDS start site falls within $w$ bp downstream of the inverton end coordinate, taking strand orientation into account. Let $T_{\mathrm{obs}}(w)$ denote the observed hit rate, defined as the fraction of invertons with at least one downstream CDS start within the specified distance. To generate an appropriate null model, we preserved both the host contig and the length of each inverton. Specifically, for each inverton $i$ with length $L_i$ on contig $c(i)$, we randomly sampled a replacement interval of length $L_i$ uniformly along $c(i)$, ensuring the interval remains within contig boundaries. This procedure was repeated for $B$ replicates to generate a null distribution of hit rates $\{T_b(w)\}_{b=1}^B$ for each window size evaluated.\par

\begin{figure}[!t]
  \centering
  \includegraphics[width=0.65\textwidth]{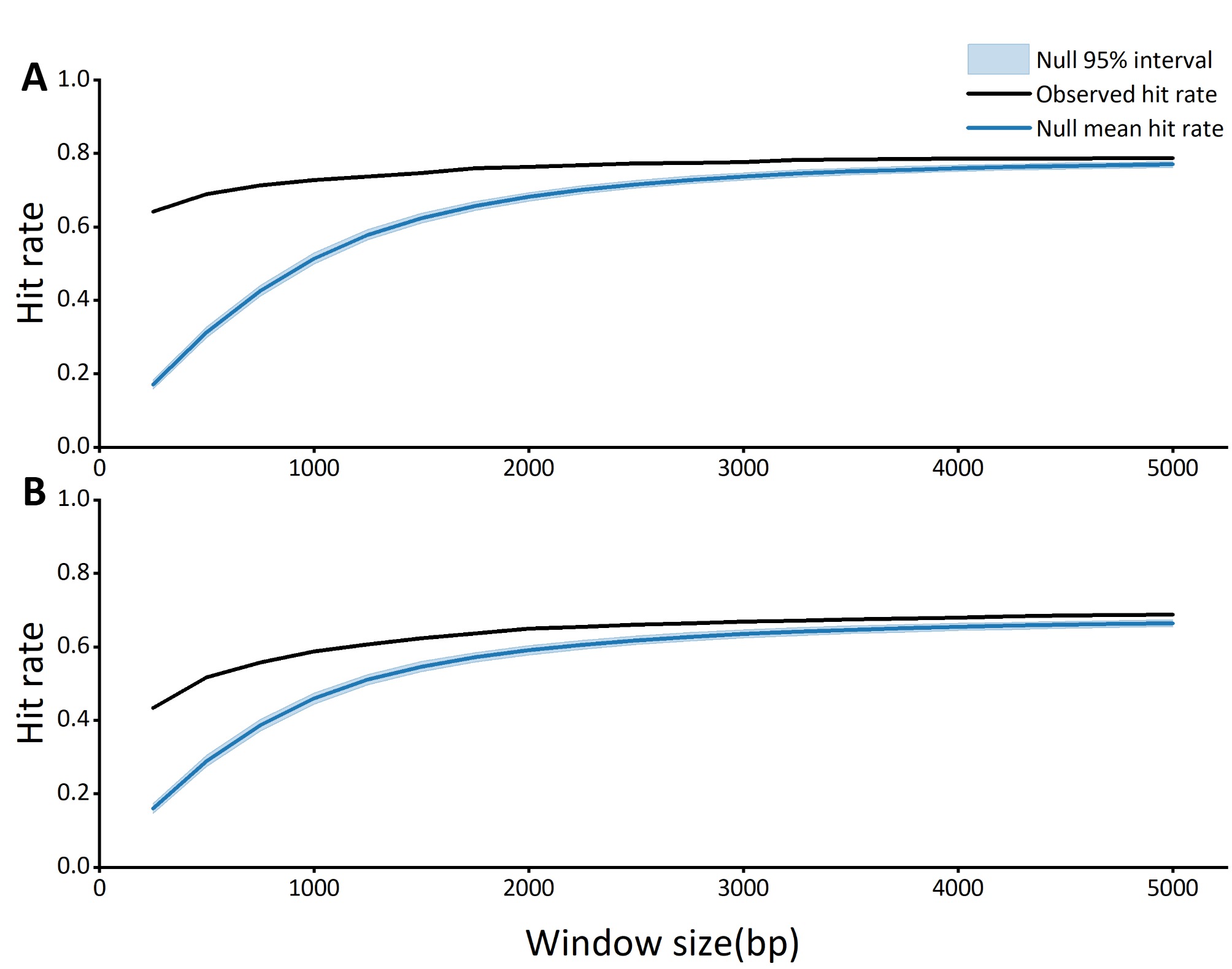}
  \caption{Enrichment of downstream coding sequences near identified invertons.
    The observed hit rate ($T_{\mathrm{obs}}$, black line) is compared to a null distribution generated by contig-preserving permutation. The dark blue line represents the mean of the null distribution, and the light blue shaded area indicates the 95\% confidence interval.
    (A) Analysis of the human gut dataset. 
    (B) Analysis of the rat gut dataset.
    In both datasets, the observed proximity of invertons to downstream CDS start sites significantly exceeds the null expectation, particularly at small window sizes.}
  \label{Fig3}
\end{figure}

Figure \ref{Fig3}A (human gut) and Figure \ref{Fig3}B (rat gut) display the observed hit rates compared against the null expectation across a continuous range of window sizes. Crucially, the observed hit rate consistently exceeds the null expectation in both datasets, a trend that is particularly pronounced at small window sizes. The gap between the observed data and the null model is widest at the shortest distances, indicating that invertons are preferentially located in the immediate vicinity of downstream coding sequences. While both the observed and null rates naturally increase as the window expands—reflecting the cumulative probability of encountering a gene over larger distances—the distinct enrichment observed at short distances suggests a tight physical linkage and potential regulatory relationship between the identified invertons and their adjacent genes.
\subsection{Putative Invertons Identified in Viral Genomes}
\begin{figure*}[!t]
  \centering
  \includegraphics[width=1.0\textwidth]{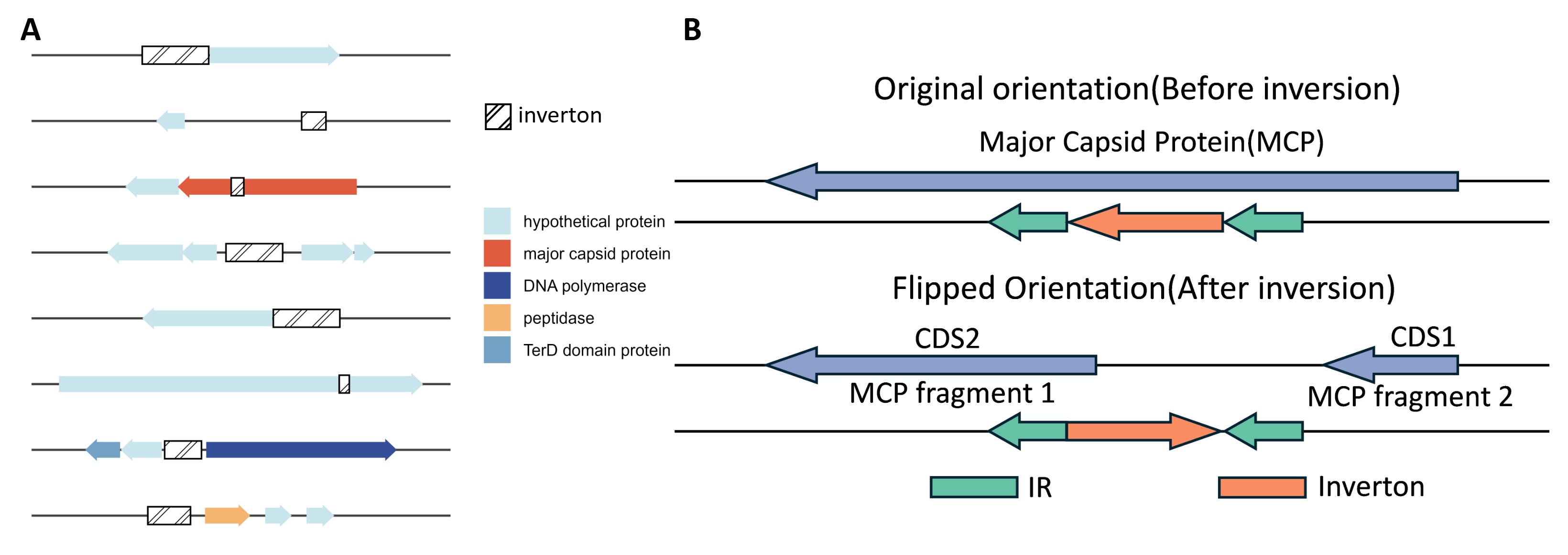}
  \caption{Genomic context and impact of identified invertons. (A) Schematic representation of coding sequences (CDSs) located within 1,000 bp downstream of or directly spanning the inverton region. Genes are color-coded by predicted function, including hypothetical proteins, major capsid protein (MCP), DNA polymerase, peptidase, and TerD domain-containing proteins. White rectangles with diagonal stripes indicate inverton sequences. (B) Representative example of a Microvirus inverton impacting the major capsid protein (MCP) gene}
  \label{Fig5}
\end{figure*}
To date, invertible DNA segments (invertons) have not been widely reported in viral genomes, despite their well-documented presence in bacteria. As our initial analyses did not distinguish between bacterial and viral origins, we subsequently performed a systematic classification to identify viral sequences. Given the high viral diversity and complex environmental pressures characteristic of the human gut, we selected the human gut dataset for this investigation. We classified contigs using four independent viral identification tools: DeepVirFinder \cite{Ren2020DeepVirFinder}, VirSorter2 \cite{Roux2015VirSorter}, ViraLM \cite{10.1093/bioinformatics/btae704}, and geNomad \cite{Camargo2024}. To ensure high specificity, we retained only those contigs supported by at least three tools, resulting in 11,415 high-confidence viral contigs. Genome quality was further assessed using CheckV \cite{Nayfach2021}. By restricting our analysis to contigs with at least ``Medium'' completeness, we obtained a final set of 1,403 quality-filtered viral contigs. Within this set, we identified putative invertons on 16 contigs, comprising 4 complete, 2 high-quality, and 10 medium-quality genomes. \par

To examine the genomic context and potential functional relevance of these candidate invertons, we predicted CDSs on the 16 contigs using Prodigal and annotated protein functions via BLASTP. The results reveal that 8 of the 16 candidate invertons are located within 1,000bp of confirmed viral CDS start or fall directly within CDS boundaries (not host-derived) (Figure \ref{Fig5}A).
These patterns suggest that, analogous to bacterial invertons which modulate gene expression or protein structure, putative viral invertons may contribute to regulatory or structural variation. Due to the absence of metatranscriptomic data, we could not directly assess expression changes of flanking genes. Therefore, to illustrate potential functional impacts, we examined a representative candidate inverton on contig ``SRR2145362\_contg9143'', which falls within a CDS boundary. This contig, classified as complete, contains a predicted invertible segment defined by inverted repeats at positions 2,653–2,666 and 2,775–2,788. The invertible interval lies entirely within a single CDS (2,083–3,993). BLASTN analysis revealed extensive alignment to \textit{Microviridae} reference sequences, and functional annotation identified the CDS as a major capsid protein (MCP). To evaluate the potential consequences of inversion, we computationally generated the reverse-complement of the invertible segment and re-annotated the modified contig. Notably, the original single MCP CDS was predicted as two distinct CDSs after inversion (Figure \ref{Fig5}B). This observation aligns with reports that Microviridae capsid proteins contain conserved modular segments \cite{annurev:/content/journals/10.1146/annurev-virology-100120-011239}, suggesting that inversion at this locus could reconfigure the MCP coding structure. Given that \textit{Microviridae} are single-stranded DNA viruses, we hypothesize that this recombination event likely occurs on the double-stranded replicative form during viral replication. Furthermore, we extended our analysis to another human gut dataset and observed similar results, which are detailed in Supplementary Section 6.
\section{Conclusion and Discussion}
In this study, we introduce TPMM, a three-component posterior mixture model designed to improve inverton detection under low-depth metagenomes. Building on the read evidence, TPMM returns posterior probabilities as soft labels, and further enables hard calling under BFDR control, reducing reliance on fixed thresholds when evidence is sparse. Validation on two gut metagenomic datasets demonstrates that TPMM recovers more invertons as read counts decrease, demonstrating practical utility in low-depth settings. Furthermore, invertons detected by TPMM exhibit clear enrichment in downstream genes, supporting their biological relevance. Notably, we observe that the majority of invertons in the human gut dataset originate from the Crohn's disease group. While this distribution may be influenced by sample imbalance, it suggests that an elevated number of invertons can potentially serve as a disease biomarker in future studies with balanced sample sizes. Moreover, we propose that viral genomes may harbor inverton sequences and provide supporting analyses and evidence. To our knowledge, this is the first work to explicitly raise and systematically investigate this possibility, expanding the potential scope of inversion mediated regulation. \par
TPMM also has limitations that motivate future improvements. First, parameter estimation may be biased when candidate numbers are small. Second, TPMM remains dependent on read evidence and cannot address candidates with no reverse read support. For such cases, TPMM cannot produce a reliable detection output. Future work may incorporate stronger regularization or hierarchical modeling to improve robustness with small candidate sets, and explore integration with sequence-level signals to enable more comprehensive inference.
\section*{Funding}
This work is supported by the Hong Kong Research Grants Council (RGC) General Research Fund (GRF) [11209823], the City University of Hong Kong projects [9667256, 9678241, 7020092] and Institute of Digital Medicine.
\section*{Acknowledgement}
We thank Prof. Fuyong Li and his student Yi Yang from Zhejiang University for the discussion about the problem formulation.
\bibliographystyle{unsrt}  
\bibliography{references}
\end{document}